\documentclass{article}

\usepackage{arxiv}
\usepackage[utf8]{inputenc} 
\usepackage[T1]{fontenc}    
\usepackage[colorlinks,citecolor=blue,urlcolor=red]{hyperref}
\usepackage{url}            
\usepackage{booktabs}       
\usepackage{amsfonts}       
\usepackage{nicefrac}       
\usepackage{microtype}      
\usepackage{graphicx}
\usepackage{natbib}
\usepackage{doi}
\usepackage{amsmath}
\usepackage{subcaption}
\usepackage{tabularx}
\usepackage{soul}

\usepackage{color}

\newcommand{\magenta}{\color{magenta}}
\usepackage{comment}

\def\mybox#1{{\magenta \vskip1mm \begin{center}
 \hspace{.0\textwidth}\vbox{\hrule\hbox{\vrule\kern6pt
  \parbox{.8\textwidth}{\kern6pt#1\vskip6pt}\kern6pt\vrule}\hrule}
  \end{center} \vskip-5mm}}

\title{A Novel Stratified Analysis Method for Testing and Estimating Overall Treatment Effects on Time-to-Event Outcomes Using Average Hazard with Survival Weight}

\date{} 					

\author{ 
    {Zihan~Qian} \\ 
	Department of Biostatistics \\
    Harvard Chan School of Public Health \\ 
    Boston, Massachusetts 02115, USA 
	\And
    {Lu~Tian}\\
	Department of Biomedical Data Science\\
    Stanford University\\ 
    Stanford, CA 94305, USA \\
	\And
    {Miki~Horiguchi}\\
    Department of Medical Oncology\\
	Dana Farber Cancer Institute\\
	Boston, MA 02215, USA \\
	\And
    {Hajime~Uno}\thanks{Email: huno@ds.dfci.harvard.edu} \\
	Department of Data Science\\
	Dana Farber Cancer Institute\\
	Boston, Massachusetts 02215, USA \\
}

\renewcommand{\headeright}{Qian et al.}
\renewcommand{\undertitle}{}
\renewcommand{\shorttitle}{Stratified analysis with Average Hazard}

\hypersetup{
pdftitle={AHstratified1},
pdfsubject={stat.app},
pdfauthor={Qian et al},
pdfkeywords={AH, Stratified},
}

\begin{document}
\maketitle

\baselineskip=12pt

\begin{abstract}
Given the limitations of using the Cox hazard ratio to summarize the magnitude of the treatment effect, alternative measures that do not have these limitations are gaining attention. 
One of the recently proposed alternative methods uses the average hazard with survival weight (AH). This population quantity can be interpreted as the average intensity of the event occurrence in a given time window that
does not involve study-specific censoring. 
Inference procedures for the ratio of AH and difference in AH have already been proposed in simple randomized controlled trial settings to compare two groups. 
However, methods with stratification factors have not been well discussed, although stratified analysis is often used in practice to adjust for confounding factors and increase the power to detect a between-group difference.  
The conventional stratified analysis or meta-analysis approach, which integrates stratum-specific treatment effects using an optimal weight, directly applies to the ratio of AH and difference in AH. However, this conventional approach has significant limitations similar to the Cochran-Mantel-Haenszel method for a binary outcome and the stratified Cox procedure for a time-to-event outcome.
To address this, we propose a new stratified analysis method for AH using standardization. With the proposed method, one can summarize the between-group treatment effect in both absolute difference and relative terms, adjusting for stratification factors. This can be a valuable alternative to the traditional stratified Cox procedure to estimate and report the magnitude of the treatment effect on time-to-event outcomes using hazard.
\end{abstract}



\section{Introduction}

Stratified analysis plays an essential role in clinical studies that compare interventions by enabling the identification of heterogeneous treatment effects, improving precision, and controlling for confounders.  Where there are known prognostic factors, these factors are used for stratified randomization, and stratified analysis will often be used as the primary analysis in randomized studies. 

When the analysis variable is a time-to-event outcome in randomized trials, the conventional analytical approach is to use the stratified log-rank test or a test based on stratified Cox regression for statistical comparison between treatment groups and to use the hazard ratio (HR) derived from the stratified Cox regression to summarize the magnitude of the treatment effect. 
However, as discussed elsewhere, this conventional stratified analysis based on Cox's HR has several limitations in providing a robust and interpretable estimate of the magnitude of the treatment effect in a target population of interest \citep{Tian2019-xw}.

The first limitation is related to its strong model assumption. The stratified Cox regression model requires that the proportional hazards (PH) assumption is maintained in each stratum. The ratio of hazard functions from two treatment groups must be constant over time in each stratum. If the PH assumption is invalid, the estimated Cox's HR depends on the underlying study-specific censoring time distribution. Therefore, interpreting or generalizing the results to future populations is problematic \citep{Kalbfleisch1981-xo,Lin:1989vg,Uno:2014ii,Horiguchi2019-xc}.

The second limitation is the lack of group-specific absolute hazards, which, while statistically advantageous, poses practical challenges. The clinical utility of the intervention cannot be evaluated solely by the HR because it is also affected by the absolute hazard in the control group. Providing absolute hazards for groups along with the HR is essential to aid clinical assessment. Some guidelines, such as CONSORT 2010 \citep{Cobos-Carbo2011-wz}, recommend reporting both absolute differences and relative effects for binary outcomes to facilitate clinical interpretation. A similar recommendation for time-to-event outcomes can be found in the guidance for authors provided by the Annals of Internal Medicine (\citeauthor{annals}, 2024). However, converting HR from the Cox's PH model into absolute hazard differences is often challenging, barring specific cases. 

The third limitation concerns the assumption of a homogeneous effect across strata, a critical consideration when employing stratified analysis methods \citep{Tian2019-xw,Sun2021-fv}. Many stratified analysis methods, including the well-known Cochran-Mantel-Haenszel method \citep{rothman2008modern}, are built on the assumption that each stratum estimates the same underlying quantity or effect. This assumption essentially posits that the magnitude of the treatment effect (e.g., HR and odds ratio) remains consistent across all strata. This assumption simplifies the analytical process, making it more manageable and interpretable. However, this homogeneity assumption is not always tenable, posing a significant limitation in interpreting the estimated magnitude of the treatment effect. 
If the homogeneity assumption is violated, these stratified analysis methods will produce a weighted average of the stratum-specific treatment effects, which may be considered a measure of the overall treatment effect in some populations. 
However, because the weights for different strata can be complex, it is unclear whether the weighted average represents the overall treatment effect in a population of interest. 

Recently, \citet{Tian2019-xw} introduced a marginal treatment effect method to overcome the third limitation associated with conventional stratified analysis techniques when dealing with various types of outcomes. 
For survival data, their method, however, is not directly applicable to Cox's HR because the hazard function is not a probabilistic measure.  Moreover, the first and second limitations concerning Cox's HR were found to be more general, rather than specific to stratified analysis. To address these complexities in the context of survival data, \citet{Tian2019-xw} proposed utilizing their method with alternative summary measures, including median survival time, cumulative incidence probability, and restricted mean survival time at specific time points.

In this paper, we further expand on the approach of \citet{Tian2019-xw} by incorporating it with an additional summary measure of event time distribution: the average hazard with survival weight (AH) \citep{Uno2023-sm}.
This measure is defined as 
\begin{equation*} 
\eta(\tau) 
= \frac{\int_{0}^{\tau} h(u)S(u)du}{\int_{0}^{\tau} S(u)du} 
= \frac{E\{I(T\le \tau)\}}{E\{ T \wedge \tau \}},
\label{AH}
\end{equation*}
where $h(u)$ and $S(u)$ are the hazard function and survival function for the event time $T,$ respectively, 
$I(A)$ is an indicator function for the event $A,$
and $x \wedge y $ denotes $\min(x,y).$
This can be interpreted as the average person-time incidence rate of $T$ on $t \in [0,\tau]$ when all $T$ before $\tau$ would have been observed without being censored by study-specific censoring time. 
The stratified analysis method we propose in this paper can provide overall group-specific AHs, the overall ratio of average hazard (RAH), and the overall difference in average hazard (DAH). This will be an additional powerful tool for analyzing time-to-event data in a more robust and informative manner than the conventional stratified Cox approach.

The structure of the rest of the paper is as follows. 
We first review and apply the conventional stratified analysis approach to AH, then propose using the marginal treatment effect approach to AH and inference procedures (Section 2). 
We apply the proposed method to data from a recently conducted cancer clinical trial (Section 3). 
We conduct numerical studies to confirm the performance of the proposed approach in finite sample size situations (Section 4) because the theoretical justification of the proposed method is based on large sample theories.
Remarks and conclusions are given in the end (Section 5).

\section{Methods}
\subsection{Notations}
Throughout this paper, we will use the following notations. 
Let $T_{jk}$ be a continuous non-negative random variable to denote the event time for group $j \ (j=0,1)$ and stratum $k \ (k=1,\ldots,K).$
Let $C_{jk}$ denote the censoring time for group $j$ and stratum $k.$ 
Assume that $T_{jk}$ is independent of $C_{jk}$ for $j=0,1,$ and $k=1,\ldots,K.$ 
Let $\left\{ (T_{jki},C_{jki}); \ i=1,\ldots,n_{jk} \right\}$ 
denote independent copies from $(T_{jk}, C_{jk}),$
where $n_{jk}$ is the sample size in group $j$ in the stratum $k.$
Let 
$X_{jki} = \min(T_{jki}, C_{jki})$ and $\Delta_{jki}=I(T_{jki} \le C_{jki}),$
where $I(A)$ is the indicator function for event $A.$ 
The observable data from group $j$ are then denoted by 
$\left\{ (X_{jki},\Delta_{jki}); \ i=1,\ldots, n_{jk}, k=1,\ldots, K  \right\}.$ 
We assume that $K$ is finite, and 
$p_{jk} = \lim_{n \rightarrow \infty} n_{jk}/n > 0 $ 
for $j=0,1,$ and $k=1,\ldots,K, $
where $ n = \sum_{j=0,1}\sum_{k=1}^{K} n_{jk}.$ 

\subsection{Conventional approaches}

First, we review the conventional stratified analysis approaches for
the treatment group comparison. 
Let $\hat{\theta}=\left(\hat{\theta}_{1},\ldots\hat{\theta}_{K}\right)^{\prime}$
be the vector of estimated treatment effects (e.g., log odds ratio
or log hazard ratio) for the $K$ strata. Let $\hat{V}(\hat{\theta})=\left(\hat{V}(\hat{\theta}_{1}),\ldots,\hat{V}(\hat{\theta}_{K})\right)^{\prime}$
be the vector that consists of the corresponding variance estimate
for $\hat{\theta}=\left(\hat{\theta}_{1},\ldots\hat{\theta}_{K}\right)^{\prime}$.
The conventional stratified analysis approach assumes that the treatment effect is the same across all strata, i.e., $\theta_{1}=\cdots=\theta_{K},$
and estimates the treatment effect by the weighted average of
$\left(\hat{\theta}_{1},\ldots\hat{\theta}_{K}\right)^{\prime}$
using the reciprocal of the variance as weight \citep{Woolf1955-br},
\begin{equation*}
\sum_{k=1}^{K}\left(\hat{\theta}_{k}/\hat{V}(\hat{\theta}_{k})\right)
/\sum_{k=1}^{K}\left(1/\hat{V}(\hat{\theta}_{k})\right)
\label{conventional_est}
\end{equation*}
and the corresponding variance estimate is given by \begin{equation*}
\left(\sum_{k=1}^{K}\hat{V}(\hat{\theta}_{k})^{-1}\right)^{-1}.
\label{conventional_var}
\end{equation*}

As \citet{Uno2023-sm} presented with the inference procedures of two-sample comparison using the ratio of AH (RAH) and the difference in AH (DAH), it is straightforward to apply this conventional approach to these between-group contrast measures. Specifically, the estimated logarithm of the ratio of AH for the stratum $k$ is given by 
\[
\hat{\theta}_{k}=\log\left({\hat{\eta}_{1k}(\tau)}/{\hat{\eta}_{0k}(\tau)}\right),
\]
 where $\hat{\eta}_{jk}(\tau)=\left\{ 1-\hat{S}_{jk}(\tau)\right\} /\hat{R}_{jk}(\tau)$
is an estimate of AH for group $j$ and stratum $k,$ and $\hat{R}_{jk}(\tau)=\int_{0}^{\tau}\hat{S}_{jk}(u) du$,
where $\hat{S}_{jk}(\cdot)$ is the Kaplan-Meier estimator for the
survival function of the event time $T_{jk}$ in the group $j$ in
the stratum $k.$ 
An approximated variance estimate of $\hat{\theta}_{k}$
is given by $\hat{V}(\hat{\theta}_{k})=\left\{ n_{1k}^{-1}\hat{V}_{1k}+n_{0k}^{-1}\hat{V}_{0k}\right\},$
where 
\begin{equation*}
\hat{V}_{jk}=\int_{0}^{\tau}\left\{ \frac{1}{1-\hat{S}_{jk}(\tau)}-\frac{\hat{R}_{jk}(u)}{\hat{R}_{jk}(\tau)}\right\} ^{2}\frac{d\hat{H}_{jk}(u)}{\hat{G}_{jk}(u)},\label{eqn:V_khat-1}
\end{equation*}
 $\hat{H}_{jk}(\cdot)$ is the Nelson-Aalen estimator for the cumulative
hazard function for group $j$ and stratum $k,$ and $\hat{G}_{jk}(t)=n_{jk}^{-1}\sum_{i=1}^{n_{jk}}I\{ \min(T_{jki}, C_{jki})\ge t \}.$
Although applying this conventional approach to the RAH and DAH is straightforward, this application shares the third limitation with the stratified Cox analysis, as discussed in the Introduction. 

One may also consider the CMH-type approach \citep{rothman2008modern}. 
This approach obtains 
a weighted average of AH's across strata for each group
and calculates DAH and RAH. 
For example, one may use the following
weights 
\[
w_{k}^{CMH1}=\frac{n_{1k}n_{0k}\hat{R}_{1k}(\tau)\hat{R}_{0k}(\tau)}{n_{1k}+n_{0k}},
\]
for $k=1,\ldots,K.$ An adjusted AH for the group $j$ is then given
by 
\[
\hat{\eta}_{j}^{CMH1}(\tau)=\left[\sum_{k=1}^{K}w_{k}^{CMH1}\left\{ \frac{1-\hat{S}_{jk}(\tau)}{\hat{R}_{jk}(\tau)}\right\} \right]/\left(\sum_{k=1}^{K}w_{k}^{CMH1}\right).
\]
The corresponding estimators for $DAH(\tau)$ and $RAH(\tau)$ are
given by 
\[
\hat{DAH}(\tau)^{CMH1}=\hat{\eta}_{1}^{CMH1}(\tau)-\hat{\eta}_{0}^{CMH1}(\tau)
\]
 and 
\[
\hat{RAH}(\tau)^{CMH1}=\hat{\eta}_{1}^{CMH1}(\tau)/\hat{\eta}_{0}^{CMH1}(\tau),
\]
respectively. 
Because $\hat{RAH}(\tau)^{CMH1}$ can be written as 
\begin{equation*}
\begin{split}
\frac{\hat{\eta}_{1}^{CMH1}(\tau)}{\hat{\eta}_{0}^{CMH1}(\tau)}
& = \frac{\sum_{k=1}^{K}\left\{ 1-\hat{S}_{1k}(\tau)\right\}\hat{R}_{0k}(\tau) (n_{0k}^{-1}+n_{{1k}}^{-1})^{-1}}{\sum_{k=1}^{K}\left\{ 1-\hat{S}_{0k}(\tau)\right\} \hat{R}_{1k}(\tau)(n_{0k}^{-1}+n_{{1k}}^{-1})^{-1}} \\
& = \left[\sum_{k=1}^{K}\left\{ \frac{\hat{\eta}_{1k}(\tau)}{\hat{\eta}_{0k}(\tau)}\right\} w_{k}^{CMH1^{*}}\right]/\left(\sum_{k=1}^{K}w_{k}^{CMH1^{*}}\right),
\end{split}
\end{equation*}
where
\[
w_{k}^{CMH1^{*}}=\frac{n_{0k}n_{1k}\left\{ 1-\hat{S}_{0k}(\tau)\right\} \hat{R}_{1k}(\tau) }{n_{1k}+n_{0k}},
\]
$\hat{RAH}(\tau)^{CMH1}$ is also a weighted average of the stratum-specific
RAH with the weights $w_{k}^{CMH1^{*}}$ for $k=1,\ldots,K.$

Since the AH is interpreted as the general censoring-free person-time
incidence rate \citep{Uno2023-sm}, one may also use the following weights
\[
w_{k}^{CMH2}=\frac{n_{1k}n_{0k}\hat{R}_{1k}(\tau) \hat{R}_{0k}(\tau) }{n_{1k}\hat{R}_{1k}(\tau) +n_{0k}\hat{R}_{0k}(\tau) },
\]
instead of $w_{k}^{CMH1}$ \citep{rothman2008modern}. The weights
$w_{k}^{CMH2}$ are interpreted as the harmonic mean of the estimated 
total observation times from two groups on the time window from 0
to $\tau$ in stratum $k$ when the effect of censoring is removed.  
$\hat{RAH}(\tau)^{CMH2}$ is then given
by 
\begin{equation*}
\begin{split}
\hat{RAH}(\tau)^{CMH2} 
& =\hat{\eta}_{1}^{CMH2}(\tau)/\hat{\eta}_{0}^{CMH2}(\tau) \\
& =\frac{\sum_{k=1}^{K}\left\{ 1-\hat{S}_{1k}(\tau)\right\}\hat{R}_{0k}(\tau)\left[\left(n_{0k}^{-1}\hat{R}_{1k}(\tau) +n_{1k}^{-1}\hat{R}_{0k}(\tau) \right)^{-1}\right]}{\sum_{k=1}^{K}\left\{ 1-\hat{S}_{0k}(\tau)\right\}\hat{R}_{1k}(\tau)\left[\left(n_{0k}^{-1}\hat{R}_{1k}(\tau) +n_{1k}^{-1}\hat{R}_{0k}(\tau) \right)^{-1}\right]} \\
& =\left[\sum_{k=1}^{K}\left\{ \frac{\hat{\eta}_{1k}(\tau)}{\hat{\eta}_{0k}(\tau)}\right\} w_{k}^{CMH2^{*}}\right]/\left(\sum_{k=1}^{K}w_{k}^{CMH2^{*}}\right),
\end{split}
\end{equation*}
where
\[
w_{k}^{CMH2^{*}}=\frac{n_{0k}n_{1k}\left\{ 1-\hat{S}_{0k}(\tau)\right\} \hat{R}_{1k}(u)du }{n_{1k}\hat{R}_{1k}(\tau) +n_{0k}\hat{R}_{0k}(\tau) }.
\]
This CMH-type approach can address the limitations of the Cox's stratified
analysis we discussed in the Introduction to some extent. However, the weighting schemes are fairly complicated and the resulting estimator is difficult to interpret, especially when the RAH is not constant across strata.

In this paper, we propose a different approach that allows users to choose any set of weights by applying the idea of standardization \citep{rothman2008modern,Tian2019-xw}. 
The proposed stratified analysis using AH addresses all the limitations of the traditional stratified Cox analysis for survival data we discussed in the previous section; it provides a robust summary of the overall treatment effect along with information that can meet the guideline recommendations for clinical trials.

\subsection{Proposed approach}
As discussed in the Introduction, providing the summary measures from two groups that yield a between-group contrast summary, such as difference or ratio, is practically important for increasing the likelihood that the between-group contrast quantity is correctly interpreted (\citeauthor{Cobos-Carbo2011-wz}, 2011;  \citeauthor{annals}, 2024). 
To address this practical need, the proposed stratified approach calculates the standardized AH for each group, adjusting for the distribution of stratification factors to a target population, and then constructs between-group contrast summaries for the magnitude of treatment effects, such as RAH and DAH. 

Because the AH is not a probability measure, we cannot obtain the standardized AH using a simple weighted average of the stratum-specific AHs. However, because the AH is a function of the survival function, we propose to define the standardized AH based on the standardized survival curve, $\bar{S}_j(t)=\sum_{k=1}^{K}w_{k} S_{jk}(t)$, 
with a given set of weights $\{w_{1,}\ldots,w_{K}\},$ that satisfies $\sum_{k=1}^{K}w_{k}=1,$ for $K$ strata. The selection of these weights should reflect the stratum sizes in the intended patient population, in which the treatment effect is studied. With a set of weights, the standardized AH is  
\begin{equation}
\bar{\eta}_{j}(\tau)
=
\frac
{ 1 - \bar{S}_j(\tau)}
{\int_{0}^{\tau} \bar{S}_j(t) du}
=
\frac
{ 1 - \left\{ \sum_{k=1}^{K}w_{k} S_{jk}(\tau) \right\} }
{\int_{0}^{\tau} \left\{ \sum_{k=1}^{K}w_{k} S_{jk}(u)\right\} du}
=
\frac
{\sum_{k=1}^{K}w_{k}\left\{ 1-S_{jk}(\tau)\right\} }{\sum_{k=1}^{K}w_{k} R_{jk}(\tau)}.
\label{standardizedAH}
\end{equation}
As a reference, in Appendix A, 
we show another argument to derive the proposed standardized AH, using the standardization method for the conventional person-time incidence rate \citep{rothman2008modern}.

Let $\hat{S}_{jk}(\cdot)$ be the Kaplan-Meier estimate of the survival function for group $j$
and stratum $k.$ 
We will estimate $\bar{\eta}_{j}(\tau)$ by 
\[
\hat{\bar{\eta}}_{j}(\tau)=\frac{\sum_{k=1}^{K}w_{k}\left\{ 1-\hat{S}_{jk}(\tau)\right\} }{\sum_{k=1}^{K}w_{k} \hat{R}_{jk}(\tau)}.
\]
In Appendix B, 
assuming that $K$ is finite and 
$p_{jk} = \lim_{n_j \rightarrow \infty} {n_{jk}}/{n_j}$ is greater than 0 for $k=1,\ldots,K,$ and 
$n_j = \sum_{k=1}^{K} n_{jk},$
we show that
$\hat{\bar{\eta}}_{j}(\tau)$ converges in probability to 
${\bar{\eta}}_{j}(\tau),$ and
also 
$Q_{j}=n_{j}^{1/2}\left\{ \hat{\bar{\eta}}_{j}(\tau)-\bar{\eta}_{j}(\tau)\right\} $
and $W_{j}=n_{j}^{1/2}\left\{ \log\hat{\bar{\eta}}_{j}(\tau)-\log\bar{\eta}_{j}(\tau)\right\}$
converge in distribution to zero-mean normal distributions with variances 
\begin{equation*}
V(Q_{j})=\sum_{k=1}^{K}\left(\frac{w_{k}^{2}}{p_{jk}}\right)
\int_{0}^{\tau}\left\{ \left(\sum_{k=1}^{K}R_{jk}(\tau)w_{k}\right)^{-1}-\left(\sum_{k=1}^{K}F_{jk}(\tau)w_{k}\right){R}_{jk}(u)\left(\sum_{k=1}^{K}R_{jk}(\tau)w_{k}\right)^{-2}\right\} ^{2}\frac{d{H}_{jk}(u)}{{G}_{jk}(u)},
\end{equation*}
and
\begin{equation*}
V(W_{j})=\sum_{k=1}^{K}\left(\frac{w_{k}^{2}}{p_{jk}}\right)
\int_{0}^{\tau}\left\{ \left(\sum_{k=1}^{K}F_{jk}(\tau)w_{k}\right)^{-1}-\left(\sum_{k=1}^{K}R_{jk}(\tau)w_{k}\right)^{-1}{R}_{jk}(u)\right\} ^{2}\frac{d{H}_{jk}(u)}{{G}_{jk}(u)},
\end{equation*}
respectively,
where 
$H_{jk}(t)$ is the cumulative hazard function for $T_{jk},$
$F_{jk}(t) = 1- S_{jk}(t),$ and
$G_{jk}(t)= \Pr(T_{jk} \wedge C_{jk}\ge t).$

We can provide the standardized AH for both groups and the corresponding confidence intervals (CIs) from these results. Specifically, an
$(1-\alpha)$ asymptotic CI for $\bar{\eta}_{j}(\tau)$ is given by
\[
\left\{ \hat{\bar{\eta}}_{j}(\tau)-z_{1-\alpha/2}\sqrt{n_{j}^{-1}\widehat{V}(Q_{j})},\hat{\bar{\eta}}_{j}(\tau)+z_{1-\alpha/2}\sqrt{n_{j}^{-1}\widehat{V}(Q_{j})}\right\} 
\]
 or 
\[
\left[\hat{\bar{\eta}}_{j}(\tau)\exp\left\{-z_{1-\alpha/2}\sqrt{n_{j}^{-1}\widehat{V}(W_{j})}\right\} ,\hat{\bar{\eta}}_{j}(\tau)\exp\left\{ z_{1-\alpha/2}\sqrt{n_{j}^{-1}\widehat{V}(W_{j})}\right\} \right],
\]
where
\begin{equation*}
\widehat{V}(Q_{j})=\sum_{k=1}^{K}\left(\frac{w_{k}^{2}}{\hat{p}_{jk}}\right)
\int_{0}^{\tau}\left\{ \left(\sum_{k=1}^{K}\hat{R}_{jk}(\tau)w_{k}\right)^{-1}-\left(\sum_{k=1}^{K}\{1-\hat{S}_{jk}(\tau)\}w_{k}\right)\hat{R}_{jk}(u)\left(\sum_{k=1}^{K}\hat{R}_{jk}(\tau)w_{k}\right)^{-2}\right\} ^{2}\frac{d\left[-\log\left\{\hat{S}_{jk}(u)\right\}\right]}{\hat{G}_{jk}(u)},
\end{equation*}
and
\begin{equation*}
\hat{V}(W_{j})=\sum_{k=1}^{K}\left(\frac{w_{k}^{2}}{\hat{p}_{jk}}\right)
\int_{0}^{\tau}\left\{ \left(\sum_{k=1}^{K}\{1-\hat{S}_{jk}(\tau)\}w_{k}\right)^{-1}-\left(\sum_{k=1}^{K}\hat{R}_{jk}(\tau)w_{k}\right)^{-1}\hat{R}_{jk}(u)\right\} ^{2}\frac{d\left[-\log\left\{\hat{S}_{jk}(u)\right\}\right]}{\hat{G}_{jk}(u)},
\end{equation*}
are consistent estimators of $V(Q_j)$ and $V(W_j),$ respectively, and $\hat{p}_{jk}$ and $\hat{G}_{jk}(u)$ are the empirical counterparts of ${p}_{jk}$ and $G_{jk}(u),$ respectively.

The DAH and RAH are then given by 
$\hat{\bar{\eta}}_{1}(\tau)-\hat{\bar{\eta}}_{0}(\tau)$
and $\hat{\bar{\eta}}_{1}(\tau)/\hat{\bar{\eta}}_{0}(\tau),$ respectively.
The corresponding $(1-\alpha)$ asymptotic CIs for $\bar{\eta}_{1}(\tau)-\bar{\eta}_{0}(\tau)$
and $\bar{\eta}_{1}(\tau)/\bar{\eta}_{0}(\tau)$ are
\[
\hat{\bar{\eta}}_{1}(\tau)-\hat{\bar{\eta}}_{0}(\tau)\pm z_{1-\alpha/2}\sqrt{n_{1}^{-1}\widehat{V}(Q_{1})+n_{0}^{-1}\widehat{V}(Q_{0})},
\]
 and
\[
\frac{\hat{\bar{\eta}}_{1}(\tau)}{\hat{\bar{\eta}}_{0}(\tau)}\exp\left[ \pm z_{1-\alpha/2}\sqrt{n_{1}^{-1}\widehat{V}(W_{1})+n_{0}^{-1}\widehat{V}(W_{0})}\right],
\]
respectively. 
For testing the null hypothesis $\bar{\eta}_{1}(\tau)=\bar{\eta}_{0}(\tau),$
either 
\[
\left\{ \hat{\bar{\eta}}_{1}(\tau)-\hat{\bar{\eta}}_{0}(\tau)\right\} /
\sqrt{n_{1}^{-1}{\widehat{V}}(Q_{1})+n_{0}^{-1}{\widehat{V}}(Q_{0})}
\]
or 
\[
\log\left\{ \frac{\hat{\bar{\eta}}_{1}(\tau)}{\hat{\bar{\eta}}_{0}(\tau)}\right\} /
\sqrt{n_{1}^{-1}{\widehat{V}}(W_{1})+n_{0}^{-1}{\widehat{V}}(W_{0})}
\]
can be used as a test statistic, each of which asymptotically follows
the standard normal distribution under the null hypothesis.

\section{Example}

\begin{figure}
\centering

\begin{subfigure}{\textwidth}
  \includegraphics[width=\linewidth]{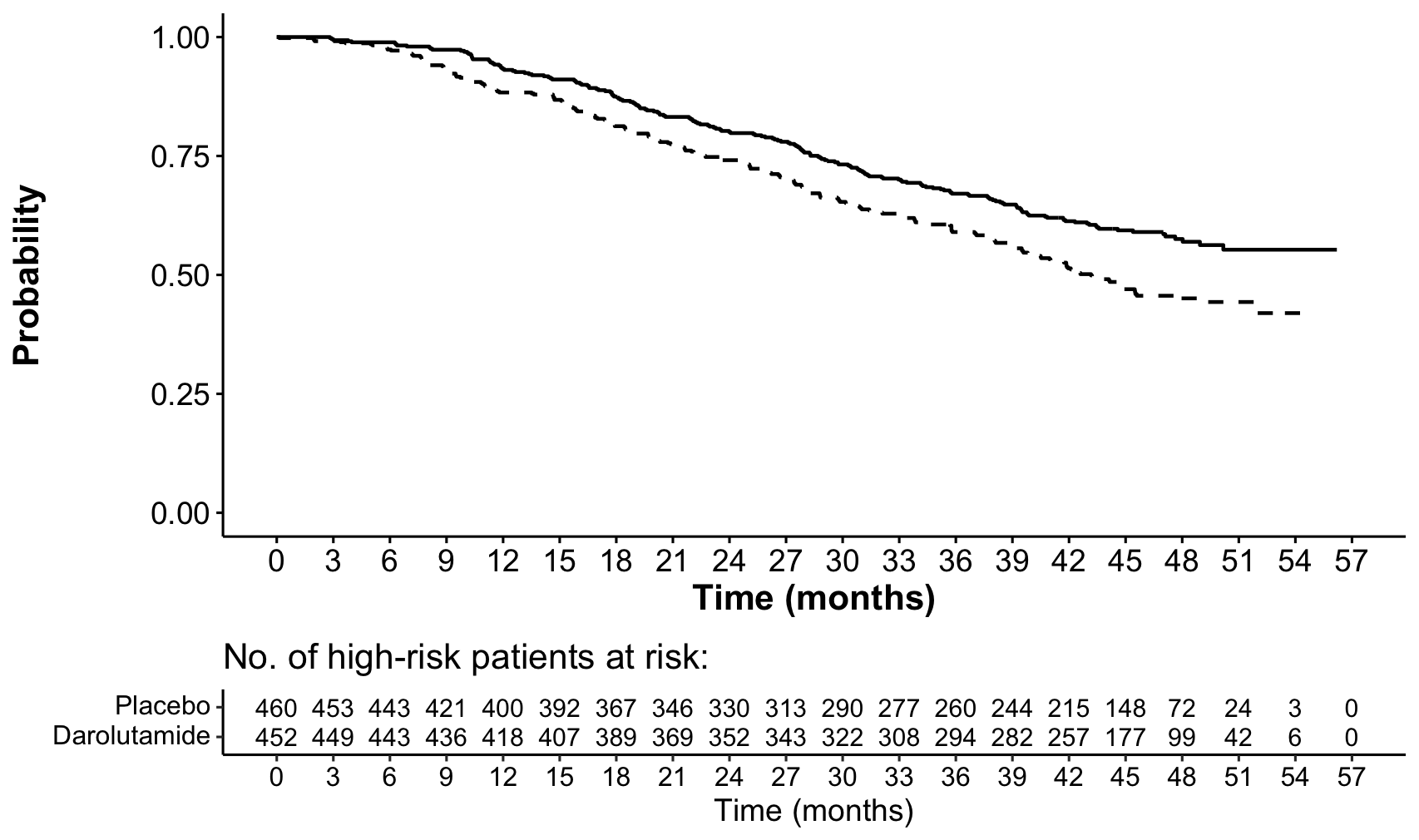}
  \captionsetup{justification=justified,singlelinecheck=false,font=normalsize}
  \caption{High-risk disease}
  \label{fig:figure1A}
\end{subfigure}

\vspace{3pt}

\begin{subfigure}{\textwidth}
  \includegraphics[width=\linewidth]{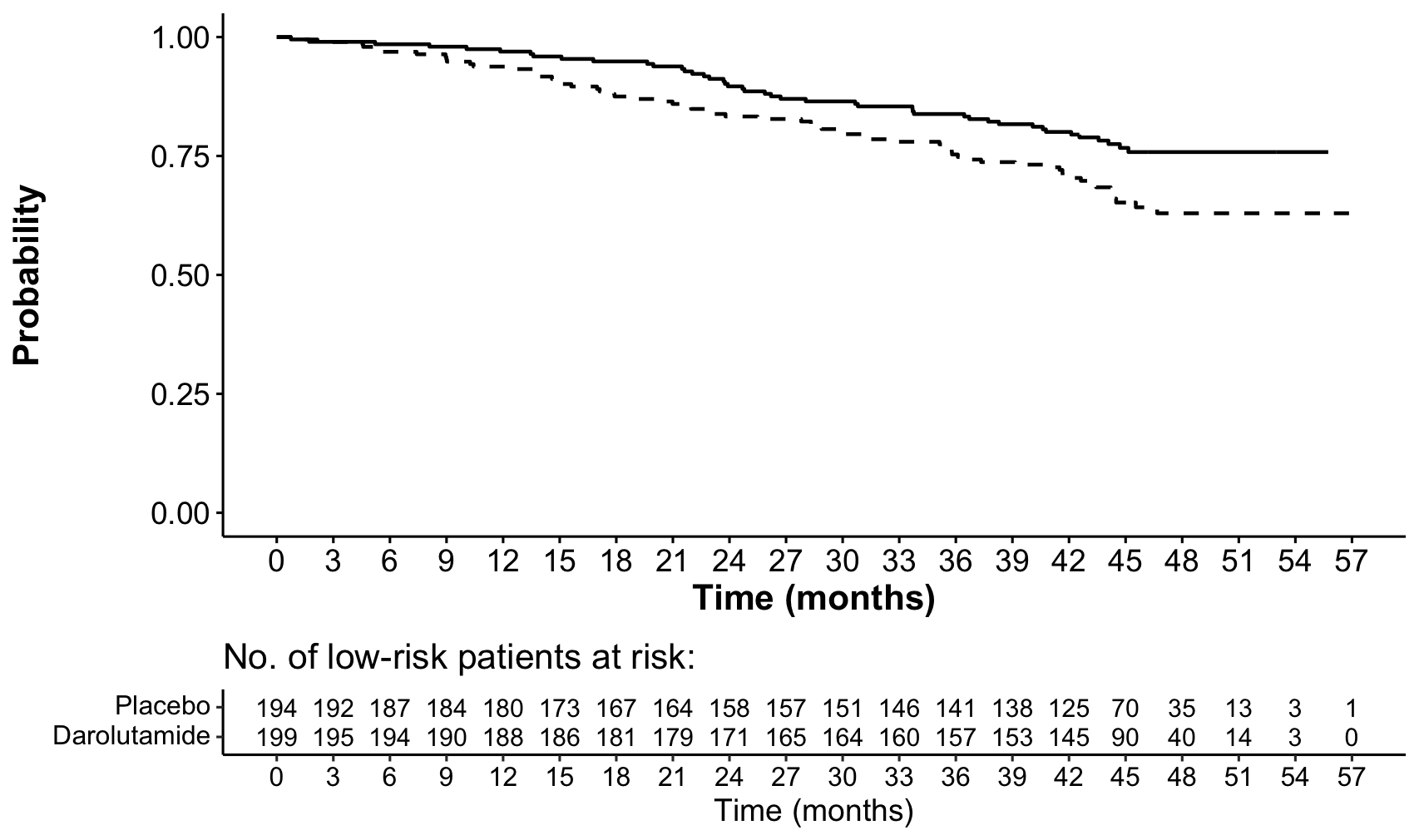}
  \captionsetup{justification=justified,singlelinecheck=false,font=normalsize}
  \caption{Low-risk disease}
  \label{fig:figure1B}
\end{subfigure}

\captionsetup{justification=justified,singlelinecheck=false}
\caption{Overall survival probabilities among patient subgroups treated with darolutamide (solid line) versus placebo (dashed line) in the ARASENS trial, stratified by disease risk: (a) High-risk and (b) Low-risk. High-risk disease was defined by two of the following three risk factors: Gleason score $\geq$ 8, $\geq$ 3 bone lesions, and measurable visceral metastases.}
\label{fig:figure1}
\end{figure}

To illustrate the proposed stratified analysis method, we use data from a randomized, phase III clinical trial comparing darolutamide plus androgen-deprivation therapy and docetaxel (darolutamide group) against placebo plus androgen-deprivation therapy and docetaxel (placebo group) in patients diagnosed with metastatic hormone-sensitive prostate cancer (ARASENS) \citep{hussain2023darolutamide}. 
In this study, the primary endpoint was overall survival (OS). Risk disease status was considered a prognostic factor and was used as a stratification factor in randomization and data analysis. High-risk disease was defined by two of the following three risk factors: Gleason score $\geq$ 8, $\geq$ 3 bone lesions, and measurable visceral metastases.
\citet{hussain2023darolutamide} also reported the Kaplan-Meier curves of OS by treatment group, dividing the patients into two strata: Stratum A consisted of patients with high-risk disease (n=912) and Stratum B consisted of patients with low-risk disease (n=393). 
We reconstructed patient-level data from the reported Kaplan-Meier curves by using the method proposed by \citet{guyot2012enhanced}. 
Figure \ref{fig:figure1} shows the Kaplan-Meier curves for OS with the reconstructed data.

First, we performed the stratified Cox analysis, which has been used routinely for stratified analysis of survival outcomes. 
The estimated HR was 0.71 (95\% CI: 0.58 to 0.86) in stratum A and 0.62 (95\% CI: 0.42 to 0.91) in stratum B,
and the estimated stratified Cox's HR was 0.69 (95\% CI: 0.58 to 0.82,  p<0.001). 
The significant p-value obtained from this conventional analysis may support a qualitative conclusion that darolutamide improves survival compared to placebo in the overall study population. 
However, interpreting the resulting quantitative information (i.e., HR=0.69) remains challenging.
The crucial assumptions on the stratified Cox analysis are 1) the PH assumption in each stratum and 2) a common HR across all strata. 
With this data, the PH assumption seems plausible in each stratum. However, 
the stratum-specific HR estimates (i.e., 0.71 in stratum A and 0.62 in stratum B) suggested a violation of the common HR. In fact, while HR = 0.69 is a weighted average of these stratum-specific HRs, the weights are not necessarily proportional to the sample size of the strata. 
Because the weights used in this approach are usually not included in the report explicitly, the interpretation of such a weighted average remains challenging in this case.    
Coupled with the lack of an absolute hazard in each treatment group, this conventional stratified Cox analysis may not provide useful quantitative information on the magnitude of the treatment effect in the overall study population. 

Next, we evaluated the treatment effect using the AH-based approach. 
We estimated the AH for each group and the absolute differences and relative ratios of the AH, accompanied by their corresponding p-values and 95\% CI. Taking into account the number of at-risk patients in each group in each stratum, we chose $\tau=48$ months as the truncation time to calculate the AH. 
Table \ref{table1} summarizes the results of the AH-based analyses.
In Stratum A, the AH was 1.102 events per 100 person-months in the darolutamide group and 1.548 in the placebo group. The corresponding DAH and RAH were -0.446 (95\% CI: -0.687 to -0.205) and 0.712 (95\% CI: 0.592 to 0.856), respectively. 
In Stratum B, the AH was 0.562 events per 100 person-months in the darolutamide group and 0.920 in the placebo group. The DAH and RAH were -0.358  (95\% CI: -0.630 to -0.086) and 0.611 (95\% CI: 0.420 to 0.888), respectively. 
Integrating these stratum-specific DAHs and RAHs, we then performed the conventional stratified analysis \citep{Woolf1955-br} and obtained -0.407 (95\% CI: -0.588 to -0.227) for DAH and 0.691 (95\% CI: 0.586 to 0.815) for RAH, both with p-values < 0.001. 
However, because the DAH and RAH in one stratum were different from those in another, the assumption of a common treatment effect across all strata was not met with the data from this study. 
Similarly to the HR estimated from the stratified Cox analysis, DAH (or RAH) derived from the conventional stratified analysis approach appeared to represent a weighted average of the two stratum-specific DAHs (or RAHs). The weights used for integrating the stratum-specific results are determined through the variances of the estimated stratum-specific DAH (or RAH), which adds complexity to the interpretation of the result. 
Furthermore, this conventional stratified analysis approach \citep{Woolf1955-br} lacks information on AH in the control and treatment groups, preventing a thorough interpretation of the magnitude of the treatment effect.

In our proposed stratified analysis using AH, we determined the weights for stratum A and stratum B as $w_1=912/(912+393)$ and $w_2=393/(912+393),$ respectively, based on the total sample size of each stratum in the analysis population. 
The adjusted AH was 0.927 (95\% CI: 0.806 to 1.047) events per 100 person-months for the darolutamide group and 1.342 (95\% CI: 1.195 to 1.489) for the placebo group. 
The corresponding adjusted DAH was -0.415 (95\% CI: -0.605 to -0.225; p < 0.001), and the adjusted RAH was 0.690 (95\% CI: 0.583 to 0.818; p < 0.001). 
Compared to the conventional stratified Cox analysis, the proposed stratified analysis method provides more robust, reliable, and richer information about the magnitude of the treatment effect on time-to-event outcomes, which will be more beneficial for treatment decision-making based on {\it hazard}. 

\begin{table}
\centering
\caption{Average hazards with survival weight with truncation time $\tau=48$ months based on separate analysis by risk disease status, conventional stratified analysis, and proposed stratified analysis for the darolutamide group and placebo group with the data reconstructed from the publication by \citet{hussain2023darolutamide} }

\scalebox{0.83}{
\label{table1}
\begin{tabular}{lcccc}
  \hline
 & Darolutamide (0.95CI) & Placebo (0.95CI) & Difference (0.95CI; p-value) & Ratio (0.95CI; p-value) \\ 
  \hline
  Stratum A (N=912) & 1.102 (-2.274 to 4.479) & 1.548 (-2.562 to 5.659) & -0.446 (-0.687 to -0.205; <0.001) & 0.712 (0.592 to 0.856; <0.001)\\
  Stratum B (N=393) & 0.562 (-1.622 to 2.745) & 0.920 (-1.892 to 3.731) & -0.358 (-0.630 to -0.086; $~~~$0.010) & 0.611 (0.420 to 0.888; $~~~$0.010)\\ 
  Conventional & - & - & -0.407 (-0.588 to -0.227; <0.001) & 0.691 (0.586 to 0.815; <0.001)\\ 
  Proposed & 0.927 (0.806 to 1.047) & 1.342 (1.195 to 1.489) & -0.415 (-0.605 to -0.225; <0.001) & 0.690 (0.583 to 0.818; <0.001)\\ 
  \hline
\end{tabular}
}

\centering
\vspace{0.3cm}
\begin{minipage}{480pt}
Note: The unit of the average hazard was expressed as the number of events per 100 person-months. \\
Abbreviations: Stratum A, high-risk disease; Stratum B, low-risk disease; CI, confidence interval; Darolutamide, darolutamide plus androgen-deprivation therapy and docetaxel group; Placebo, placebo plus androgen-deprivation therapy and docetaxel group.\\
\end{minipage}
\end{table}

\section{Numerical Studies}
\subsection*{4.1 Configurations}

While asymptotic theories justify the proposed stratified analysis method we proposed in Section 2, the statistical properties are still unclear when the sample size is finite in practical situations. We conducted numerical studies to address this issue and evaluated the statistical performance of the proposed nonparametric inference procedure. 

In these numerical studies, we utilized the reconstructed patient-level data we analyzed in the previous section to mimic the metastatic hormone-sensitive prostate cancer study reported by  \citet{hussain2023darolutamide}.
Specifically, we fit a Weibull model and calculated the maximum likelihood estimates (MLEs) of the Weibull distribution parameters for each of the four subgroups delineated by the treatment groups 
(darolutamide and placebo) and strata (A and B). 
The MLEs of shape and scale parameters for each of the four subgroups were as follows:
1.52 and 69.62 (darolutamide group in stratum A),
1.46 and 55.87 (placebo group in stratum A),
1.43 and 118.65 (darolutamide group in stratum B), and
1.37 and 87.64 (placebo group in stratum B).
These estimated Weibull distributions generated the event time data throughout the numerical studies. 

We considered two patterns regarding censoring: (I) random censoring that is common across all four subgroups and (II) no random censoring. The specification of the censoring pattern (I) was also based on the prostate cancer data. We fit a Weibull distribution for censoring time with the data from the four subgroups pooled together. The estimated shape and scale parameters were 8.21 and 47.79, respectively.

Regarding the sample size, we considered two scenarios: n=700 and n=1400 per arm. 
The allocation of sample sizes to the four subgroups was determined by mimicking the prostate cancer data \citep{hussain2023darolutamide}.
Since, in the prostate cancer data, the sample sizes for the darolutamide and placebo groups were 452 and 460, respectively, in stratum A, and 199 and 194, respectively, in stratum B, we allocated 70\% of the sample size to stratum A and 30\% to stratum B. Within each stratum, we allocated 50\% to each of the two treatment groups. 

While we used $\tau= 48$ months as the truncation time point for the AH calculation in the previous section, here we considered three different \(\tau\) values, 45,  48, and 51 months, in the numerical studies. As such, we simulated a total of 12 configurations. 

For each $\tau,$ we calculated the true AH values using the estimated Weibull distributions for the four subgroups. 
First, the true AH for group $j$ was computed by utilizing the formula \( \eta_j(\tau)=\sum_{k=1}^2 w_k F_{jk}(\tau) / 
\left\{ \sum_{k=1}^2 w_k R_{jk}(\tau) \right\} \), where $w_1=0.7,$ $w_2=0.3,$ \( F_{jk}(\tau) \) is the cumulative incidence probability at \(\tau\) in group $j$ and stratum $k,$ and \(R_{jk}(\tau)=\int_0^\tau \{1-F_{jk}(u)\}du\) is the restricted mean survival time at \(\tau\). The true DAH and RAH were then derived as $\eta_1(\tau) - \eta_0(\tau)$ and $\eta_1(\tau) / \eta_0(\tau),$ respectively.

For a given sample size, we generated the observable data for the group $j$ and stratum $k,$ represented as \(X_{jk} = \min(T_{jk}, C_{jk})\) and \(\Delta_{jk} = I(T_{jk} \leq C_{jk})\), where \(T_{jk}\) is the event time and \(C_{jk}\) is the censoring time.
We then applied the proposed stratified analysis to the observable data. We estimated the adjusted AH values for the darolutamide and placebo groups and corresponding DAH and log(RAH) along with 95\% CIs for the three different \(\tau\) values. 
This entire simulation process was repeated 3,000 times, and bias of the point estimate  
and coverage probability of the corresponding 95\% CI were calculated by comparing the results of each simulation data with the true values. 
We also calculated the minimum average size of the risk sets at each time point of $\tau$ across two strata for each treatment group to investigate the relationship between these sizes and the statistical performance.

\subsection*{4.2 Results}
Table \ref{table2} shows the simulation results for the censoring pattern (I: random censoring).
We confirmed that the empirical biases were negligibly small and the coverage probabilities were sufficiently close to the nominal level of 0.95 in all settings for the AHs of the darolutamide and placebo groups, DAH, and log(RAH) in both sample sizes n=700 and n=1400. 
We observed similar findings with the scenarios under the censoring pattern (II: no random censoring) (see Table \ref{table3}). 
In summary, the proposed stratified analysis method demonstrates promising performance in terms of accuracy of point estimates and achieving accurate coverage probabilities.
This performance is promising across various sample sizes and $\tau$ values as long as the size of risk set at $\tau$ is large enough to support the reliance of asymptotic theories generally.  This underscores the potential applicability of our stratified approach in practice.


\begin{table}
\caption{Bias and coverage probability of the proposed stratified analysis method with various \(\tau\) based on 3000 iterations under the censoring pattern (I: random censoring).}
\label{table2}
\centering
\scalebox{0.9}{
\begin{tabular}{lcccccc}
\hline
\multicolumn{1}{l}{} & \multicolumn{3}{c}{N = 700} & \multicolumn{3}{c}{N = 1400} \\
\cmidrule(lr){2-4} \cmidrule(lr){5-7}
\(\tau\) & 45 & 48 & 51 & 45 & 48 & 51 \\
\hline
True Value\(^1\)  & & & & & & \\
AH darolutamide & 0.911 & 0.935 & 0.958 & 0.911 & 0.935 & 0.958\\
AH placebo & 1.303 & 1.331 & 1.357 & 1.303 & 1.331 & 1.357 \\
DAH & -0.393 & -0.396 & -0.399 &  -0.393 & -0.396 & -0.399 \\
log(RAH) & -0.359 & -0.353 & -0.348 & -0.359 & -0.353 & -0.348 \\
\hline
min (n)\(^2\) & & & & & & \\
AH darolutamide & 80.4 & 50.0 & 24.7 & 159.5 & 99.1 & 48.5 \\
AH placebo & 39.2 & 24.9 & 12.6 & 77.3 & 49.0 & 24.5 \\
\hline
Bias\(^1\) & & & & & & \\
AH darolutamide & -0.002 & -0.002 & -0.002 & 0.001 & 0.001 & 0.001 \\
AH placebo & 0.002 & 0.002 & 0.002 & 0.001 & 0.000 & 0.001 \\
DAH & -0.004 & -0.004 & -0.004 & 0.001 & 0.001 & 0.000 \\
log(RAH) & -0.005 & -0.005 & -0.005 & 0.000 & 0.001 & 0.000 \\
\hline
Coverage\(^3\) & & & & & & \\
AH darolutamide & 0.961 & 0.959 & 0.957 & 0.955 & 0.958 & 0.952 \\
AH placebo & 0.958 & 0.956 & 0.958 & 0.955 & 0.959 & 0.960 \\
DAH & 0.959 & 0.961 & 0.954 & 0.956 & 0.952 & 0.959 \\
log(RAH) & 0.959 & 0.963 & 0.959 & 0.957 & 0.952 & 0.958 \\
\hline
\end{tabular}
}
\centering

\vspace{0.3cm}
\begin{minipage}{270pt}
\textit{Note}: 
\( ^1 \)The unit of AH is expressed as the number of events per 100 person-months; 
\( ^2 \)The minimum of the average sizes of risk set at \(\tau\) from two strata; 
\( ^3 \)The nominal coverage probability is 95\%.
\end{minipage}
\end{table}

\begin{table}
\caption{Bias and coverage probability of the proposed stratified analysis method with various \(\tau\) based on 3000 iterations under the censoring pattern (II: no random censoring).}
\label{table3}
\centering
\scalebox{0.9}{
\begin{tabular}{lcccccc}
\hline
\multicolumn{1}{l}{} & \multicolumn{3}{c}{N=700} & \multicolumn{3}{c}{N=1400} \\
\cmidrule(lr){2-4} \cmidrule(lr){5-7}
\(\tau\) & 45 & 48 & 51 & 45 & 48 & 51 \\
\hline
True Value\(^1\)   & & & & & & \\
AH darolutamide & 0.911 & 0.935 & 0.958 & 0.911 & 0.935 & 0.958 \\
AH placebo & 1.303 & 1.331 & 1.357 & 1.303 & 1.331 & 1.357 \\
DAH & -0.393 & -0.396 & -0.399 & -0.393 & -0.396 & -0.399 \\
log(RAH) & -0.359 & -0.353 & -0.348 & -0.359 & -0.353 & -0.348 \\
\hline
min (n)\(^2\) & & & & & &\\
AH darolutamide & 146.9 & 139.4 & 132.0 & 293.3 & 278.1 & 263.3 \\
AH placebo & 71.5 & 69.0 & 66.4 & 141.5 & 136.3 & 131.3 \\
\hline
Bias\(^1\)  & & & & & &\\
AH darolutamide & 0.001 & 0.001 & 0.001 & 0.001 & 0.001 & 0.001 \\
AH placebo & 0.001 & 0.000 & 0.000 &0.000 & 0.001 & 0.001 \\
DAH & 0.001 & 0.001 & 0.001 & 0.001 & 0.000 & 0.000 \\
log(RAH) & 0.000 & 0.001 & 0.001 & 0.000 & 0.000 & 0.000 \\
\hline
Coverage\(^3\) & & & & & & \\
AH darolutamide & 0.962 & 0.963 & 0.966 & 0.953 & 0.954 & 0.952 \\
AH placebo & 0.956 & 0.952 & 0.956 & 0.957 & 0.957 & 0.962 \\
DAH & 0.961 & 0.961 & 0.959 & 0.951 & 0.953 & 0.954 \\
log(RAH) & 0.963 & 0.965 & 0.962 &0.951 & 0.955 & 0.955 \\
\hline
\end{tabular}
}
\centering

\vspace{0.3cm}
\begin{minipage}{270pt}
\textit{Note}: 
\( ^1 \)The unit of AH is expressed as number of events per 100 person-months; 
\( ^2 \)The minimum of the average sizes of risk set at \(\tau\) from two strata; 
\( ^3 \)The nominal coverage probability is 95\%.
\end{minipage}
\end{table}

\section{Remarks}
In this paper, we proposed a new stratified analysis procedure for the AH, applying the idea of direct standardization. 
The proposed method allows users to estimate the adjusted treatment effect in terms of both difference and relative terms (i.e., DAH and RAH, respectively) following the recommendation of reporting the estimated magnitude of the treatment effect (\citeauthor{Cobos-Carbo2011-wz}, 2011;  \citeauthor{annals}, 2024). 
Unlike the conventional stratified Cox analysis, the proposed method does not require the proportional hazards assumption within each stratum or a common hazard ratio across all strata.
We believe the proposed method is a powerful stratified analysis tool to present the treatment effect on time-to-event outcome in a more robust and informative manner, effectively overcoming the limitations of the conventional stratified Cox approach.

In this paper, we assumed a finite number of strata, denoted as $K$. Since the justification of the proposed inference procedure is based on large sample theories, each stratum must have a sufficiently large sample size to ensure the accuracy of the large-sample approximation. If some strata have small sample sizes, merging them is a practical solution to enhance stability. If one is interested in conducting a stratified analysis with a large number of small strata, weighting each individual with a stratum-specific weight could be an alternative solution.

The software for the implementation of the proposed method is currently
R and is included in the latest version of the \textit{survAH2} package,
which is available upon request from the corresponding author. 

\section*{Appendix} 
\appendix
\section{The standardization of the conventional person-time incidence rate and the standardized AH}
As described in Section 1, the AH can be interpreted as an average person-time incidence rate on a time window $[0,\tau],$ with the effect of censoring before 
$\tau$ removed. 
We can apply the standardization method for the conventional person-time incidence rate to the AH. 

The standardized method for the conventional person-time incidence
rate is performed as follows \citep{rothman2008modern}.
Let $E_{k}$ and $L_{k}$ be the total number of observed events and the total exposure time in the stratum $k.$ 
The conventional person-time incidence rate is given by $$I_{k}=E_{k}/L_{k}.$$ 
Let ${W_{k},k=1,\ldots,K}$ denote a set of person-time in the reference population. 
The standardized incidence rate is then given by the weighted average of 
${I_{k},k=1,\ldots,K}$
using ${W_{k},k=1,\ldots,K}$ as weights. 
Specifically, it is 
\begin{equation}
\bar{I}=\frac{\sum_{k=1}^{K}I_{k}W_{k}}{\sum_{k=1}^{K}{W_{k}}}=\frac{\sum_{j=1}^{K}E_{k}W_{k}/L_{k}}{\sum_{k=1}^{K}{W_{k}}}\label{eq:IR}
\end{equation}

Now, we apply this standardization approach to the AH. 
By definition, let 
\begin{equation}
{\eta}_{k}(\tau)=\frac{\left\{ 1-S_{k}(\tau)\right\} }{\int_{0}^{\tau}S_{k}(u)du}=\frac{F_{k}(\tau)}
{R_{k}(\tau)}=\frac{E\{I(T_{k}\le\tau)\}}
{E\{T_{k}\wedge\tau\}}\label{eqn:eta_j}
\end{equation}
denote the AH in the stratum $k.$ Here, we multiply the sample
size of the stratum $k,$ $n_{k},$ to both the numerator and denominator of the last term in the equation (\ref{eqn:eta_j}).
The numerator is then
\[
E_{k}^{*}={n_{k}E\{I(T_{k}\le\tau)\}}=n_{k}F_{k}(\tau)
\]
and the denominator is
\[
L_{k}^{*}={n_{k}E\{T_{k}\wedge\tau\}}=n_{k}R_{k}(\tau).
\]
These are interpreted as the total number of events we would observe and the total exposure time in the stratum $k,$ respectively, when no censored observations would have been observed before $\tau.$ 

Now, we want to obtain the standardized AH from the stratum-specific AH 
$\{\eta_{k}(\tau)=E_{k}^{*}/L_{k}^{*},k=1,\ldots,K\}.$ 
Let $N_{k}$ be the sample size of the stratum $k$ in the reference population for $k=1,\ldots,K$. 
Then, the person-time for the stratum $k$ in the reference population will be 
\[
W_{k}^{*}=N_{k}R_{k}(\tau).
\]
By plugging $E_{k}^{*},L_{k}^{*}$ and $W_{k}^{*}$ into $E_{k},L_{k}$ and $W_{k}$ in equation (\ref{eq:IR}), respectively, we have the standardized AH as 
\[
\bar{I}^{*}=
\frac{\sum_{k=1}^{K}E_{k}^{*}W_{k}^{*}/L_{k}^{*}}
{\sum_{k=1}^{K}{W_{k}^{*}}}
=\frac{\sum_{k=1}^{K}n_{k}F_{k}(\tau)N_{k}R_{k}(\tau)/\left(n_{k}R_{k}(\tau)\right)}{\sum_{k=1}^{K}N_{k}R_{k}(\tau)}
=\frac{\sum_{k=1}^{K}N_{k}F_{k}(\tau)}
{\sum_{k=1}^{K}N_{k}R_{k}(\tau)}.
\]
If we replace $\{N_1,\ldots,N_K \}$ in the above equation by $\{w_1,\ldots,w_K\},$ this is identical to the standardized AH (\ref{standardizedAH}) that we define in Section 2. 

\section{Large sample properties of $Q_{j}$ and $W_{j}$}

We use the same notation and assumptions as in Section 2. Note that
$\{w_{1,}\ldots,w_{K}\}$ is a given weight set and $\sum_{k=1}^{K}w_{k}=1.$
First,
we rewrite $Q_{j}=n_{j}^{1/2}\left\{ \frac{\sum_{k=1}^{K}\hat{F}_{jk}(\tau)w_{k}}{\sum_{k=1}^{K}\hat{R}_{jk}(\tau)w_{k}}-\frac{\sum_{k=1}^{K}F_{jk}(\tau)w_{k}}{\sum_{k=1}^{K}R_{jk}(\tau)w_{k}}\right\} $
by 
\begin{eqnarray*}
Q_{j} & = & n_{j}^{1/2}\left\{ \sum_{k=1}^{K}\hat{F}_{jk}(\tau)w_{k}-\sum_{k=1}^{K}F_{jk}(\tau)w_{k}\right\} \left(\sum_{k=1}^{K}\hat{R}_{jk}(\tau)w_{k}\right)^{-1}\\
 &  & +n_{j}^{1/2}\left(\sum_{k=1}^{K}F_{jk}(\tau)w_{k}\right)\left\{ \left(\sum_{k=1}^{K}\hat{R}_{jk}(\tau)w_{k}\right)^{-1}-\left(\sum_{k=1}^{K}R_{jk}(\tau)w_{k}\right)^{-1}\right\} ,
\end{eqnarray*}
where $F_{jk}(t)=1-S_{jk}(t)$ and $R_{jk}(t)=\int_{0}^{t}S_{jk}(u)du.$
By the application of the Taylor series expansion, it can be shown that
\begin{equation*}
Q_{j}=\sum_{k=1}^{K}w_{k}p_{jk}^{-1/2}U_{jk}+o_{p}(1),
\label{eq:Qj}
\end{equation*}
where $p_{jk}=n_{jk}/n_{j}$ in the proportion of sample size in the
stratum $k$ in the group $j,$ and 
\begin{equation}
U_{jk}=n_{jk}^{1/2}\left[\left(\sum_{k=1}^{K}R_{jk}(\tau)w_{k}\right)^{-1}\left\{ \hat{F}_{jk}(\tau)-F_{jk}(\tau)\right\} -\left(\sum_{k=1}^{K}F_{jk}(\tau)w_{k}\right)\left(\sum_{k=1}^{K}R_{jk}(\tau)w_{k}\right)^{-2}\left\{ \hat{R}_{jk}(\tau)-R_{jk}(\tau)\right\} \right].
\label{eq:Ujk}
\end{equation}
On the other hand, as it is shown by \citet{FH1991},
\citet{Zhao:2012bp}, 
and \citet{Uno2023-sm}, 
\begin{equation}
\sqrt{n}_{jk}\left\{ \frac{\hat{F}_{jk}(\tau)-{F}_{jk}(\tau)}{1-F_{jk}(\tau)}\right\} =n_{jk}^{-1/2}\sum_{i=1}^{n_{jk}}\int_{0}^{\tau}\frac{dM_{jki}(u)}{G_{jk}(u)}+o_{p}(1),\label{eqn:F-dist-1}
\end{equation}
and 
\begin{equation}
\sqrt{n}_{jk}\left\{ \hat{R}_{jk}(\tau)-{R}_{jk}(\tau)\right\} =-n_{jk}^{-1/2}\sum_{i=1}^{n_{jk}}\int_{0}^{\tau}\left\{ \int_{u}^{\tau}S_{jk}(t)dt\right\} \frac{dM_{jki}(u)}{G_{jk}(u)}+o_{p}(1),\label{eqn:R-dist-1}
\end{equation}
where $G_{jk}(t)=\Pr(X_{jk}\ge t),$ $M_{jki}(t)=N_{jki}(t)-\int_{0}^{t}Y_{jki}(s)dH_{jk}(s),$
$N_{jki}(t)=I(X_{jki}\le t,\Delta_{jki}=1),$ and $Y_{jki}(t)=I(X_{jki}\ge t).$
As $n_{jk}$ goes to $\infty,$ these converge weakly to a zero-mean normal distribution.
Plugging in these (\ref{eqn:F-dist-1}) and (\ref{eqn:R-dist-1})
into the equation (\ref{eq:Ujk}), 
we have
\[
U_{jk}=n_{jk}^{-1/2}\sum_{i=1}^{n_{jk}}\int_{0}^{\tau}\left\{ \left(\sum_{k=1}^{K}R_{jk}(\tau)w_{k}\right)^{-1}-\left(\sum_{k=1}^{K}F_{jk}(\tau)w_{k}\right){R}_{jk}(u)\left(\sum_{k=1}^{K}R_{jk}(\tau)w_{k}\right)^{-2}\right\} \frac{dM_{jki}(u)}{G_{jk}(u)}+o_{p}(1).
\]
Similarly, by the martingale central limit theorem, 
as $n_{jk}$ goes to $\infty,$ 
it is shown that $U_{jk}$
converges weakly to a normal distribution with mean zero and variance
\[
{V}(U_{jk})=\int_{0}^{\tau}\left\{ \left(\sum_{k=1}^{K}R_{jk}(\tau)w_{k}\right)^{-1}-\left(\sum_{k=1}^{K}F_{jk}(\tau)w_{k}\right){R}_{jk}(u)\left(\sum_{k=1}^{K}R_{jk}(\tau)w_{k}\right)^{-2}\right\} ^{2}\frac{d{H}_{jk}(u)}{{G}_{jk}(u)}.
\]
Therefore, $Q_{j}=\sum_{k=1}^{K}w_{k}p_{jk}^{-1/2}U_{jk}$ also converges
weakly to a normal distribution with mean zero and variance
\begin{equation*}
V(Q_{j})=\sum_{k=1}^{K}\left(\frac{w_{k}^{2}}{p_{jk}}\right)V(U_{jk}).\label{eq:VarQj}
\end{equation*}

Similarly, it is straightforward to show that 
\[
W_{j}=n_{j}^{1/2}\left\{ \log\left(\frac{\sum_{k=1}^{K}\hat{F}_{jk}(\tau)w_{k}}{\sum_{k=1}^{K}\hat{R}_{jk}(\tau)w_{k}}\right)-\log\left(\frac{\sum_{k=1}^{K}F_{jk}(\tau)w_{k}}{\sum_{k=1}^{K}R_{jk}(\tau)w_{k}}\right)\right\} =\sum_{k=1}^{K}w_{k}p_{jk}^{-1/2}\xi_{jk}+o_{p}(1),
\]
converges weakly to a normal distribution with mean zero and variance
\begin{equation*}
V(W_{j})=\sum_{k=1}^{K}\left(\frac{w_{k}^{2}}{p_{jk}}\right)V(\xi_{jk}),\label{eq:VarWj}
\end{equation*}
where 
\begin{eqnarray*}
\xi_{jk} & = & n_{jk}^{1/2}\left[\left(\sum_{k=1}^{K}F_{jk}(\tau)w_{k}\right)^{-1}\left\{ \hat{F}_{jk}(\tau)-F_{jk}(\tau)\right\} -\left(\sum_{k=1}^{K}F_{jk}(\tau)w_{k}\right)^{-1}\left\{ \hat{R}_{jk}(\tau)-R_{jk}(\tau)\right\} \right]\\
 & = & n_{jk}^{-1/2}\sum_{i=1}^{n_{jk}}\int_{0}^{\tau}\left\{ \left(\sum_{k=1}^{K}R_{jk}(\tau)w_{k}\right)^{-1}-\left(\sum_{k=1}^{K}F_{jk}(\tau)w_{k}\right){R}_{jk}(u)\left(\sum_{k=1}^{K}R_{jk}(\tau)w_{k}\right)^{-2}\right\} \frac{dM_{jki}(u)}{G_{jk}(u)}+o_{p}(1),
\end{eqnarray*}
and
\begin{equation*}
{V}(\xi_{jk})=\int_{0}^{\tau}\left\{ \left(\sum_{k=1}^{K}F_{jk}(\tau)w_{k}\right)^{-1}-\left(\sum_{k=1}^{K}R_{jk}(\tau)w_{k}\right)^{-1}{R}_{jk}(u)\right\} ^{2}\frac{d{H}_{jk}(u)}{{G}_{jk}(u)}.\label{eqn:Var_xi}
\end{equation*}

\subsection*{Financial disclosure}
Research reported in this publication was supported by the National Institute of General Medical Sciences of the National Institutes of Health under award number R01GM152499 (HU, LT), the National Heart Lung and Blood Institute
 of the National Institutes of Health under award number R01HL089778 (LT), and the McGraw/Patterson Research Fund (HU).

\subsection*{Conflict of interest}
The authors declare no potential conflict of interest.

\subsection*{Data availability statement}
Data supporting the findings of this study are available from the corresponding author upon reasonable request. 

\bibliographystyle{biom}
\bibliography{refs}

\end{document}